\documentclass[aps,pra,preprint,superscriptaddress]{revtex4-1}

\usepackage{indentfirst}
\usepackage{bm}
\usepackage{graphicx}
\usepackage{epsfig}
\usepackage{amsmath}
\usepackage{amsfonts}
\usepackage{amssymb}
\usepackage{amsthm}
\usepackage{latexsym}

\usepackage{booktabs}
\usepackage{caption}
\newtheorem{theorem}{Theorem}
\newtheorem{lemma}{Lemma}

\newtheorem{corollary}{Corollary}
\usepackage{epsfig}
\usepackage{amsmath}
\usepackage{epstopdf}
\usepackage{pgf,fancyhdr}
\usepackage{float}
\usepackage[mathscr]{euscript}

\usepackage{hyperref}

\begin{document}
\title{Detection of Genuine Multipartite Entanglement in Arbitrary Multipartite systems}
\author{Yu Lu}
\email{2210502158@cnu.edu.cn}
\affiliation{School of Mathematical Sciences, Capital Normal University, 100048 Beijing, China}
\author{Shao-Ming Fei}
\email{feishm@cnu.edu.cn }
\affiliation{School of Mathematical Sciences, Capital Normal University, 100048 Beijing, China}

\bigskip

\begin{abstract}
We study the genuine multipartite entanglement of arbitrary $n$-partite quantum states by representing the density matrices in terms of the generalized Pauli operators. We introduce a general framework for detecting genuine multipartite entanglement and non full-separability of multipartite quantum states with arbitrary dimensions based on correlation tensors. Effective criterion is derived to verify the genuine multipartite entanglement.
Detailed examples are given to show that the criterion detects more genuine multipartite entanglement than the existing criteria.
\bigskip

\textbf{Keywords} Genuine entanglement, Correlation tensor, Generalized Pauli operators
\end{abstract}

\maketitle
	
\section{Introduction}
Quantum entanglement is a key resource in quantum information processing  \cite{niesen2000quantum, divincenzo1995quantum} such as entanglement swapping \cite{bose1998multiparticle}, quantum cryptography \cite{ekert1991quantum} and quantum secure communication \cite{bennett1992communication}. Towards the quantification of entanglement various entanglement measures have been proposed~\cite{bennett1996concentrating,hill1997entanglement,vidal1999entanglement,carvalho2004decoherence,emary2004relation,osterloh2005constructing}.
An important type of entanglement in multipartite quantum systems is the genuine multipartite entanglement (GME). A multipartite quantum state is said to be genuine multipartite entangled if it is not separable with respect to any bi-partitions \cite{guhne2009entanglement}. Genuine multipartite entanglement plays important roles in measurement-based quantum computation \cite{briegel2009measurement} and various quantum communication protocols \cite{De2011QuantumAI} such as secret sharing (\cite{gisin2002quantum},\cite{karlsson1999quantum}), extreme spin squeezing \cite{sorensen2001entanglement}, metrology \cite{hillery1999quantum}, quantum computing \cite{raussendorf2001one} and quantum networks \cite{hyllus2012fisher,PhysRevLett.87.117901,zhao2004experimental}. Series criteria have been derived to detect genuine multipartite entanglement \cite{huber2010detection, huber2014witnessing,ma2011measure,ma2012entanglement,bancal2014device,lancien2015relaxations}.

In this article, based on the Weyl representation with the generalized Pauli operators,
an effective criterion is derived to verify the genuine multipartite entanglement.
We present detailed examples to illustrate the advantages of our criterion, showing that
the criterion detects more genuine multipartite entanglement than the existing criteria.
	
\section{Detection of genuine multipartite entanglement}
Consider $n$-partite systems in vector space $H_{1}^{d_{1}}\otimes H_{2}^{d_{2}}\otimes ...\otimes H_{n}^{d_{n}}$ with individual dimensions $d_s$, $s=1,2,...n$.
The generalized Pauli operators on the $s$th $d_{s}$-dimensional Hilbert space $H_{s}^{d_{s}}$ are given by
\begin{equation}\label{1}
A_{\mu_{s}}^{(s)}=A_{{d_{s}i+j}}^{(s)}=\sum_{m=0}^{d_{s}-1}\omega ^{im}E_{m,m+j}, \qquad (0\le\mu_{s}\le d_{s}^{2}-1)
\end{equation}
where $\omega$ is the $d_s$-th primitive root of unity $\omega^{d_s}=1$, $E_{ij}$ is the $d_s\times d_s$ matrix with the only nonzero entry 1 at the position ${(i,j)}$,
and by means of the division with remainder $\mu\in{{0,\cdots,d_{s}^{2}-1}}$ is expressed uniquely as $\mu_{s}=d_{s}i+j$ for a pair $(i,j)$ ($0\le i,j\le d_{s}-1$).
The basis given by the generalized Pauli operators obeys the following algebraic relations,
$A_{d_si+j}^{(s)}A_{d_sk+l}^{(s)}=\omega^{jk}A_{d_s(i+k)+(j+l)}^{(s)}$, $(A_{d_si+j}^{(s)})^{\dagger}=\omega^{ij}A_{(d_s-i)+(d_s-j)}^{(s)}$ and $tr(A_{d_si+j}^{(s)}(A_{d_sk+l}^{(s)})^{\dagger})=\delta_{ik}\delta_{jl}d_{s}$.

An $n$-partite state in $H_{1}^{d_{1}}\otimes H_{2}^{d_{2}}\otimes ...\otimes H_{n}^{d_{n}}$ can be generally expressed as
\begin{equation}\label{rho}
\rho = \frac{1}{d_{1}d_{2}\cdots d_{n}}\sum_{s=1}^n\sum_{\mu_{s}=0}^{d_{s}^{2}-1}t_{\mu_{1},\mu_{2},\cdots,\mu_{n}}A_{\mu_{1}}^{(1)}\otimes A_{\mu_{2}}^{(2)}\otimes\cdots\otimes A_{\mu_{n}}^{(n)},
\end{equation}
where $A_{0}^{(s)}=I_{d_{s}}$ $(s=1,2,\cdots,n)$, $t_{\mu_{1},\mu_{2},\cdots, \mu_{n}}=tr(\rho(A_{\mu_{1}}^{(1)})^{\dagger}\otimes (A_{\mu_{2}}^{(2)})^{\dagger}\otimes\cdots\otimes (A_{\mu_{n}}^{(n)})^{\dagger})$ are complex coefficients.
Denote by $\Vert \cdot \Vert$ the norm of a (column) complex vector, i.e., $\Vert \nu \Vert = \sqrt{\nu^{\dagger}\nu}$. The trace norm of a retangular matrix $A\in \mathbb{C}^{m\times n}$ is defined by $\Vert A \Vert_{tr}=\sum \sigma_i=tr\sqrt{AA^{\dagger}}$, where $\sigma_i $ are the singular values of A. Clearly $\Vert A \Vert_{tr}=\Vert A^{\dagger} \Vert_{tr}$.
Denote ${\dagger}$ the conjugate and transpose of a matrix. We first give several lemmas that will be used in proving our main results.

\begin{lemma}
Let $M$ be an $m\times m$ matrix. Let $\lambda_{1}\ge \cdots \ge \lambda_{m}$ and $S_{1}\ge \cdots \ge S_{m}$ be the eigenvalues of $\frac{M+M^{\dagger}}{2}$ and $(MM^{\dagger})^\frac{1}{2}$, respectively. Then we have $\lambda_{i}\le S_{i}$ for $1\le i\le m$.
\end{lemma}

\begin{proof}
Denote by $|\upsilon_{1}\rangle, |\upsilon_{2}\rangle, \dots, |\upsilon_{m}\rangle$ and $|\omega_{1}\rangle, |\omega_{2}\rangle, \dots, |\omega_{m}\rangle$ the corresponding eigenvectors of $\frac{M+M^{\dagger}}{2}$ and $(MM^{\dagger})^\frac{1}{2}$, respectively, $\frac{M+M^{\dagger}}{2}|\upsilon_{i}\rangle=\lambda_{i}|\upsilon_{i}\rangle$, $(MM^{\dagger})|\omega_{i}\rangle={S_i}^2|\omega_{i}\rangle$. We have
$$
dim(span\{|\upsilon_{1}\rangle, |\upsilon_{2}\rangle, \dots, |\upsilon_{k}\rangle\}\cap span\{|\omega_{k}\rangle, \dots, |\omega_m\rangle\})\ge 1
$$
for any fixed $k$. Then there exists vector $|x\rangle$ such that
$$
\langle x|\frac{M+M{^\dagger}}{2}|x\rangle\ge\lambda_{k}~~~ and ~~~ \langle x| MM^{\dagger}|x\rangle\le (s_{k})^{2}.
$$
Therefore, $\lambda_k\le \langle x|\frac{M+M{^\dagger}}{2}|x\rangle = Re \langle x| M^\dagger |x\rangle \le |\langle x|M^\dagger|x\rangle| \le s_k$.
\end{proof}

\begin{lemma}
Let $A$ and $B$ be $m\times m$ hermitian semi-positive definite matrices.
Then there are unitary matrices $U$ and $V$ such that
\begin{equation}\label{l2}
\left[ A+B \right]\le U\left[A \right]U^{\dagger}+V \left[B\right] V^{\dagger},
\end{equation}
where $\left[M\right]=\left(MM^{\dagger}\right)^{\frac{1}{2}}$ for any matrix $M$.
\end{lemma}

\begin{proof}
Let $C=A+B=\frac{A+A^\dagger}{2}+\frac{B+B^\dagger}{2}$ as $A$ and $B$ are hermitian.
By polar decomposition we have $C=SQ$, where $S$ is semi-positive and $Q$ is unitary. $C$ is also a semi-positive definite matrix.
There are unitary matrices $U_1$ and $U_2$ such that $U_1^\dagger[\frac{A+A^\dagger}{2}]U_1=D_1$, $U_2^\dagger[\frac{A+A^\dagger}{2}]U_2=D_2$ or $U_1^\dagger U^\dagger[A]UU_1=D_2$,
where $D_{1}$ and $D_{2}$ are diagonal matrices, $U=U_2 U_1^\dagger$.
Since $\frac{A+A^\dagger}{2}$ and $U^{-1} [A]U$ can be diagonalized at the same time, we have
$\frac{A+A^\dagger}{2}\le U[A]U^\dagger$. Similarly we can obtain $\frac{B+B^\dagger}{2}\le U[B]U^\dagger$.

As $C\le U[A]U^\dagger+V[B]V^\dagger$ and $C$ is semi-positive, we have $[C]=\left(CC^{\dagger}\right)^{\frac{1}{2}}=\left(C^{2}\right)^{\frac{1}{2}}=C$.
Hence, $[C]\le U[A]U^\dagger+V[B]V^\dagger$, which proves the lemma.
\end{proof}

\begin{lemma}
Let $F=\begin{bmatrix} S_{1} \\ S_{2} \end{bmatrix}$ be a block matrix given by matrices $S_{1}$ and $S_{2}$.
Then $\begin{Vmatrix} F \end{Vmatrix}\le \begin{Vmatrix} S_1 \end{Vmatrix} + \begin{Vmatrix} S_2 \end{Vmatrix}$.
\end{lemma}
	
\begin{proof}
According to the definition we have $\begin{Vmatrix} F \end{Vmatrix} = tr(FF^\dagger)^\frac{1}{2} = tr(F^\dagger F)^\frac{1}{2} = tr(S_1^\dagger S_1 +S_2^\dagger S_2)^\frac{1}{2}$.
Denote $A=S_1^\dagger S_1$, $B=S_2^\dagger S_2$ and $C=FF^\dagger=A+B$.
According to Lemma 1, (\ref{l2}) can be also expressed as $tr(CC^\dagger)^\frac{1}{2} \le tr(U[A]U^\dagger+V[B]V^\dagger)\le tr(AA^\dagger)^\frac{1}{2}+tr(BB^\dagger)^\frac{1}{2}$.
According to \eqref{3} we have $\begin{Vmatrix} C \end{Vmatrix} \le \begin{Vmatrix} A \end{Vmatrix} + \begin{Vmatrix} B \end{Vmatrix}$. Therefore,
\begin{align*}
\begin{Vmatrix} F \end{Vmatrix}& = \sqrt{\begin{Vmatrix} FF^\dagger \end{Vmatrix}}= \sqrt{\begin{Vmatrix} C \end{Vmatrix}}\le \sqrt{\begin{Vmatrix} A \end{Vmatrix}+\begin{Vmatrix} B \end{Vmatrix}}\le \sqrt{\begin{Vmatrix} A \end{Vmatrix}} + \sqrt{\begin{Vmatrix} B \end{Vmatrix}}\\
 &= \sqrt{\begin{Vmatrix} S_1^\dagger S_1 \end{Vmatrix}} + \sqrt{\begin{Vmatrix} S_2^\dagger S_2 \end{Vmatrix}}  = \begin{Vmatrix} S_1 \end{Vmatrix} + \begin{Vmatrix} S_2 \end{Vmatrix}.
\end{align*}
\end{proof}

We now consider a general $n$-partite state $\rho\in H_1^{d_1}\otimes H_2^{d_2}\otimes\cdots\otimes H_n^{d_n}$ given by \eqref{rho}.
Suppose that $\rho$ is separable under the bipartition $l_1\cdots l_{k-1}|l_k\cdots l_{n}$.
Let $T^{(l_1\cdots l_k)}$ be the $(d^2_{l_1}-1)\cdots (d^2_{l_k}-1)$-dimensional column vector with entries $t_{u_{l_1}\cdots u_{l_k}0\cdots0}$
associated with the coefficients $\rho$ in the generalized Pauli operators representation. Define
$S_0^{l_1\cdots l_{k-1}|l_k}$ to be the block matrix
$S_0^{l_1\cdots \l_{k-1}|\l_k}=[S^{l_1\cdots \l_{k-1}|\l_k}~~O_{l_1\cdots l_{k-1}}]$, where
$S^{l_1\cdots l_{k-1} l_k}=T^{(l_1\cdots l_{k-1})}(T^{(l_k)})^{\dagger}=([t_{u_{l_1}\cdots u_{l_k}0\cdots0}])$ is the
$\prod\limits_{s=1}^{k-1}(d_{l_s}^2-1)\times (d_{l_k}^2-1)$ matrix and
$O_{l_1\cdots l_{k-1}}$ is the $\prod\limits_{s=1}^{k-1}(d_{l_s}^2-1)\times
[\prod\limits_{s=k}^{n}(d_{l_s}^2-1)-(d_{l_k}^2-1)]$
zero matrix. Define $S^{l_1\cdots l_{k-1}|l_k\cdots l_n}=T^{(l_1\cdots l_{k-1})}(T^{(l_k\cdots l_n)})^{\dagger}=[t_{u_1,\cdots, u_n}]$ to be a $\prod\limits_{s=1}^{k-1}(d_{l_s}^2-1)\times\prod\limits_{s=k}^{n}(d_{l_s}^2-1)$ matrix.

Set for real numbers $\alpha$, $\beta$ and distinct
indices $l_1, \ldots, l_n\in\{1, 2, \cdots, n\}$,
\begin{equation}\label{}
F^{l_1\cdots l_{k-1}|l_k\cdots l_{n}}=\begin{bmatrix}\alpha S_0^{l_1\cdots \l_{k-1}|l_k} \\ \beta S^{l_1\cdots l_{k-1}|l_k\cdots l_n} \end{bmatrix},
\end{equation}
where $k-1=1, 2,\cdots, [n/2]$, $[n/2]$ denotes the smallest integer less or equal to $n/2$.
For example, when $\rho\in H_1^{2}\otimes H_2^{2}\otimes H_3^{2}\otimes H_4^{3}$, we have
$$
F^{14|23}=\begin{bmatrix}\alpha S_0^{14|2} \\ \beta S^{14|23} \end{bmatrix},
$$
where
$$
S^{14|2}=\left[
\begin{array}{ccc}
t_{1,1,0,1}~~ & t_{1,2,0,1}~~ & t_{1,3,0,1} \\
t_{1,1,0,2}~~ & t_{1,2,0,2}~~ & t_{1,3,0,2} \\
t_{1,1,0,3}~~ & t_{1,2,0,3}~~ & t_{1,3,0,3} \\
\vdots & \vdots & \vdots \\
t_{3,1,0,3}~~ & t_{3,2,0,3}~~ & t_{3,3,0,3}
\end{array}
\right]
$$
and
$$
S^{14|23}=\left[
\begin{array}{cccccc}
t_{1,1,1,1}~ & t_{1,1,2,1} & \cdots & t_{1,1,8,1} & \cdots & t_{1,3,8,1}\\
t_{1,1,1,2}~ & t_{1,1,2,2} & \cdots & t_{1,1,8,2} & \cdots & t_{1,3,8,2}\\
t_{1,1,1,3}~ & t_{1,1,2,3} & \cdots  & \cdot  & \cdots  & \cdot \\
\vdots~ & \vdots & \vdots  & \vdots  & \vdots  & \vdots  \\
t_{3,1,1,3}~ & t_{3,1,2,3} & \cdots  & \cdot  & \cdots  & \cdot
\end{array}
\right].
$$

\begin{theorem}\label{3}
If an $n$-partite state $\rho\in H_1^{d_1}\otimes H_2^{d_2}\otimes\cdots\otimes H_n^{d_n}$ is separable under the bipartition $l_1\cdots l_{k-1}|l_k\cdots l_{n}$, we have that\\
	(i) $\|F^{l_1|l_{2}\cdots l_{n}}\|_{tr}\leq W_{l_1}$;\\
	(ii) $\|F^{l_1\cdots l_{k-1}|l_k\cdots l_{n}}\|_{tr}\leq W_{l_1\cdots l_{k-1}}$ $(k\geq3)$;\\
	where
	\begin{footnotesize}
		\begin{equation*}
		W_{l_1}=\sqrt{d_{l_1}-1}\left(|\alpha|\sqrt{d_{l_2}-1}+|\beta|
		\sqrt{\frac{d_{l_2}\cdots d_{l_n}(n-2-\sum\limits_{s=2}^nd_{l_s}^{-1})+1}{n-2}}\right),
		\end{equation*}
		\begin{equation*}
		W_{l_1\cdots l_{k-1}}=\sqrt{\frac{d_{l_1}\cdots d_{l_{k-1}}(k-2-\sum\limits_{s=1}^{k-1}d_{l_s}^{-1})+1}{k-2}}
		\left(|\alpha|\sqrt{d_{l_k}-1}+|\beta|\sqrt{\frac{d_{l_k}\cdots d_{l_n}(n-k-\sum\limits_{s=k}^nd_{l_s}^{-1})+1}{n-k}}\right).
		\end{equation*}
	\end{footnotesize}
\end{theorem}

\begin{proof}
We shall use repeatedly
$\Vert T^{(1)} \Vert_{tr}\leq d_{1}-1$ for $\rho\in H_{1}^{d_1}$ and
$$
\Vert T^{(12\cdots n)}\Vert^2\leq \frac{d_1\cdots d_n(n-1-\sum\limits_{s=1}^n\frac{1}{d_s^2})+1}{n-1}
$$	
for $n\geq 2$, $\rho\in H_1^{d_1}\otimes H_2^{d_2}\otimes\cdots\otimes H_n^{d_n}$ \cite{zhao2022detection}.

$(i)$ If an $n$-partite mixed state $\rho_{l_1l_2\cdots l_{n}}$ is separable under the bipartition $l_1|l_2\cdots l_{n}$, it can be expressed as
$\rho_{l_1l_2\cdots l_{n}}=\sum\limits_s p_s\rho_{l_1}^s\otimes\rho_{l_2\cdots l_{n}}^s$, $0<p_s\leq1$, $\sum\limits_s p_s=1$,
where
$$\rho_{l_1}^s=\frac{1}{d_{l_1}}\sum_{u_{l_1}=0}^{d_{l_1}^2-1} t_{u_{l_1}}^sA_{u_{l_1}}^{(l_1)}$$
and
$$\rho_{l_2\cdots l_{n}}^s=\frac{1}{d_{l_2}\cdots d_{l_{n}}}\sum\limits_{q=2}^{n}\sum_{u_{l_q}=0}^{d_{l_q}^2-1}t_{u_{l_2},\cdots,u_{l_n}}^s
A_{u_{l_2}}^{(l_2)}\otimes \cdots\otimes A_{u_{l_{n}}}^{(l_{n})}.
$$
Then
$S^{l_1|l_2}=\sum\limits_sp_sT_s^{(l_1)}(T_s^{(l_2)})^{\dagger}$ and $S^{l_1|l_2\cdots l_n}=\sum\limits_sp_sT_s^{(l_1)}(T_s^{(l_2\cdots l_n)})^{\dagger}$.
From Lemma 3 we obtain
\begin{equation*}
	\begin{split}
	\Vert F^{l_1|l_2\cdots l_{n}}\Vert_{tr}&\leq \Vert \alpha S^{l_1|l_2}\Vert+\Vert \beta S^{l_1|l_2\cdots l_n}\Vert\\ &=\sum_sp_s(|\alpha|\|T_s^{(l_1)}\|\|T_s^{(l_2)}\|+|\beta|\|T_s^{(l_1)}\|\|T_s^{(l_2\cdots l_n)}\|)\\
	&\leq\sqrt{d_{l_1}-1}\left(|\alpha|\sqrt{d_{l_2}-1}+|\beta|
	\sqrt{\frac{d_{l_2}\cdots d_{l_n}(n-2-\sum\limits_{s=2}^n\frac{1}{d_{l_{s}^{2}}})+1}{n-2}}\right)\\
	&=W_{l_1}.
	\end{split}
	\end{equation*}
	
$(ii)$ If $\rho$ is separable under the bipartition $l_1\cdots l_{k-1}|l_k\cdots l_{n}$, it can be expressed as
$\rho_{l_1\cdots l_{k-1}|l_k\cdots l_{n}}=\sum\limits_s p_s\rho_{l_1\cdots l_{k-1}}^s\otimes\rho_{l_k\cdots l_{n}}^s$ with $0<p_s\leq 1$ and $\sum\limits_s p_s=1$,
where
$$\rho_{l_1\cdots l_{k-1}}^s=\frac{1}{d_{l_1}\cdots d_{l_{k-1}}}\sum\limits_{p=1}^{k-1}\sum_{u_{l_p}=0}^{d_{l_p}^2-1}
	t_{u_{l_1},\cdots,u_{l_{k-1}}}^s
	A_{u_{l_1}}^{(l_1)}\otimes \cdots\otimes A_{u_{l_{k-1}}}^{(l_{k-1})}
$$
and
$$
\rho_{l_k\cdots l_{n}}^s=\frac{1}{d_{l_k}\cdots d_{l_{n}}}\sum\limits_{q=k}^{n}\sum_{u_{l_q}=0}^{d_{l_q}^2-1}t_{u_{l_k},\cdots,u_{l_{n}}}^s
	A_{u_{l_k}}^{(l_k)}\otimes \cdots\otimes A_{u_{l_{n}}}^{(l_{n})}.
$$
Then $S^{l_1\cdots l_{k-1}|l_k}=\sum\limits_sp_sT_s^{(l_1\cdots \l_{k-1})}(T_s^{(l_k)})^{\dagger}$ and
$S^{l_1\cdots l_{k-1}|l_k\cdots l_n}=\sum\limits_sp_sT_s^{(l_1\cdots \l_{k-1})}(T_s^{(l_k\cdots l_n)})^{\dagger}$.
In a similar way we get
	\begin{equation*}
	\begin{split}
	\ &\Vert F^{l_1\cdots l_{k-1}|l_k\cdots l_{n}}\Vert_{tr}\leq \Vert \alpha S^{l_1\cdots l_{k-1}|l_k}\Vert+\Vert \beta S^{l_1\cdots l_{k-1}|l_k\cdots l_n}\Vert\\
	&=\sum_sp_s(|\alpha|\|T_s^{(l_1\cdots l_{k-1})}\|\|T_s^{(l_k)}\|+|\beta|\|T_s^{(l_1\cdots l_{k-1})}\|\|T_s^{(l_k\cdots l_n)}\|)\\
	\leq&\sqrt{\frac{d_{l_1}\cdots d_{l_{k-1}}(k-2-\sum\limits_{s=1}^{k-1}\frac{1}{d_{l_s}^{2}})+1}{k-2}}
	[|\alpha|\sqrt{d_{l_k}-1}+|\beta|\sqrt{\frac{d_{l_k}\cdots d_{l_{n}}(n-k-\sum\limits_{s=k}^n\frac{1}{d_{l_{s}^{2}}})+1}{n-k}}]\\
	=&W_{l_1\cdots l_{k-1}}.
	\end{split}
	\end{equation*}
\end{proof}

Now we consider the detection of genuine multipartite entanglement. An $n$-partite mixed state $\rho=\sum p_i|\varphi_i\rangle\langle\varphi_i|$ is said to be biseparable if $|\varphi_i\rangle$ $(i=1,2,\cdots,n)$ can be expressed as one of the forms: $|\varphi_i\rangle=|\varphi_i^{l_1\cdots\l_{k-1}}\rangle\otimes|\varphi_i^{l_k\cdots l_n}\rangle$, where $|\varphi_i^{l_1\cdots\l_{k-1}}\rangle$ and $|\varphi_i^{l_k\cdots l_n}\rangle$ are some pure states in $H_{l_1}^{d_{l_1}}\otimes\cdots \otimes H_{l_{k-1}}^{d_{l_{k-1}}}$ and $H_{l_k}^{d_{l_k}}\otimes\cdots \otimes H_{l_{n}}^{d_{l_{n}}}$, respectively, $l_1\neq \cdots\neq l_n\in\{1,2,\cdots,n\}$. Otherwise, $\rho$ is said to be genuine multipartite entangled. Let
\begin{equation}
T(\rho)=\frac{1}{\sum\limits_{s=1}^mC_n^s}(\sum\limits _{l_1=1}^n
\Vert F^{l_1|l_2\cdots l_n}\Vert_{tr}+\sum\limits _{1\leq l_1<\atop l_2\leq n}\Vert F^{l_1l_2|l_3\cdots l_n}\Vert_{tr}+\cdots+\sum\limits _{1\leq l_1<\cdots\atop<l_m\leq n}\Vert F^{l_1\cdots l_m|l_{m+1}\cdots l_n}\Vert_{tr}),
\end{equation}
where $C_n^s=\frac{n!}{s!(n-s)!}$, $m=[n/2]$. Set $K_2=\textrm{Max}\{W_{l}, W_{l_1\cdots l_{k-1}}\}$, where $l=1,2,\cdots,n$, $l_1<\cdots<\l_{k-1}\in\{1,2,\cdots,n\}$.
We have the following criterion.

\begin{theorem}
A mixed state $\rho\in H_1^{d_1}\otimes H_2^{d_2}\otimes\cdots\otimes H_n^{d_n}$ is genuine multipartite entangled if $T(\rho)>K_2$.
\end{theorem}

\begin{proof}
If $\rho$ is biseparable, it follows from Theorem 1 that
\begin{equation*}
\begin{split}
T(\rho)&=\frac{1}{\sum\limits_{s=1}^mC_n^s}(\sum\limits _{l_1=1}^n\Vert F^{l_1|l_2\cdots l_n}\Vert_{tr}+\sum\limits _{1\leq l_1<\atop l_2\leq n}\Vert F^{l_1l_2|l_3\cdots l_n}\Vert_{tr}+\cdots+\sum\limits _{1\leq l_1<\cdots\atop<l_m\leq n}\Vert F^{l_1\cdots l_m|l_{m+1}\cdots l_n}\Vert _{tr})\\
&\leq\frac{1}{\sum\limits_{s=1}^mC_n^s}(K_2+\cdots+K_2)\leq K_2.
\end{split}
\end{equation*}
Consequently, if $T(\rho)>K_2$, $\rho$ is genuine multipartite entangled.
\end{proof}

In particular, for permutational invariant states $\rho$, i.e., $\rho=P\rho P^{\dagger}$ for any permutation $P$ of the qudits, we have the following corollary.

\begin{corollary} Let $\rho$ be a permutational invariant density matrix.
$\rho$ is genuine multipartite entangled if $T(\rho)>J_2$, where
$$
J_2=\frac{1}{\sum\limits_{s=1}^mC_n^s}(\sum\limits _{l=1}^nW_{l}+\sum\limits _{1\leq l_1< l_2\leq n}W_{l_1l_2}+\cdots+\sum\limits _{1\leq l_1<\cdots<l_m\leq n}W_{l_1\cdots l_m}).
$$
\end{corollary}

\begin{figure}[H]
    \centering
    \includegraphics[width=0.70\textwidth]{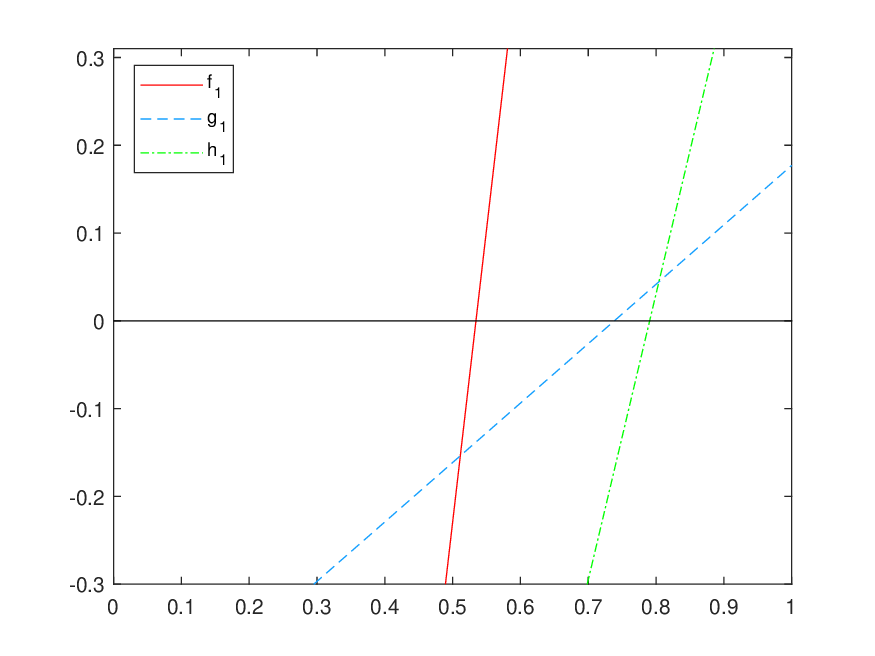}\\
    \caption{$f_1(x)$ from our result (solid red line), $g_1(x)$ from the Theorem 2 in \cite{li2017measure} (dashed blue line) and $h_1(x)$ from the Theorem 3 in \cite{de2011multipartite}(dash-dotted green line).}
    \label{fig:1}
    \end{figure}

\textit{\textbf{Example 1}} Consider the mixed three-qubit $W$ state,
\begin{equation}\label{14}
\rho_W=\frac{1-x}{8}I_8+x|W\rangle\langle W|, \quad 0\leq x\leq1,
\end{equation}
where $|W\rangle=\frac{1}{\sqrt{3}}(|001\rangle+|010\rangle+|100\rangle)$ and $I_8$ is the $8\times8$ identity matrix. Using Theorem 2 we obtain
the corresponding intervals of $x$ for which $\rho_W$ is genuine tripartite entangled with respect to different choices of $\alpha$ and $\beta$, see Table \ref{tab:1}.
In particular, for $\alpha=\frac{1}{10}$ and $\beta=2$, set $f_1(x)=T(\rho_W)-K_1=6.6688x-(\frac{1}{10}+2\sqrt{3})$. $f_1(x)>0$ gives rise to $0.5344<x\leq1$. The Theorem 2 in \cite{li2017measure} says that if $g_1(x)=\frac{1}{12}(\sqrt{66}x-6)>0$, $\rho_W$ is genuine tripartite entangled, that is, $0.7385<x\leq1$. The Theorem 2 in \cite{de2011multipartite} implies that $\rho_W$ is genuine tripartite entangled if $h_1(x)=3.26x-\frac{6+\sqrt{3}}{3}>0$, i.e., $0.791<x\leq1$. Our result clearly outperforms the above results, see Fig. \ref{fig:1}.
\begin{table}[!htb]
	\caption{$T(\rho_W)$, $K_1$ and the range of GME of the state (\ref{14}) for different $\alpha$ and $\beta$.}
	\label{tab:1}
	\centering
	\begin{tabular}{ccccc}
		\hline\noalign{\smallskip}
	    \ & ~$T(\rho_W)$ & ~$K_1$ &~ Range of GME \\
		\noalign{\smallskip}\hline\noalign{\smallskip}
		$\alpha=1,\,\beta=1$ & ~$4.7952x$ &~ $1+\sqrt{3}$ &~ $0.5697<x\leq1$ \\
		$\alpha=\frac{1}{2},\,\beta=2$ &~ $7.2704x$ & ~$\frac{1}{2}+2\sqrt{3}$ &~ $0.5452<x\leq1$ \\
		$\alpha=\frac{1}{10},\,\beta=2$ &~ $6.6688x$ & ~$\frac{1}{10}+2\sqrt{3}$ &~ $0.5344<x\leq1$\\
		\noalign{\smallskip}\hline
	\end{tabular}
\end{table}

\begin{figure}[H]
	\centering
	\includegraphics[width=0.75\textwidth]{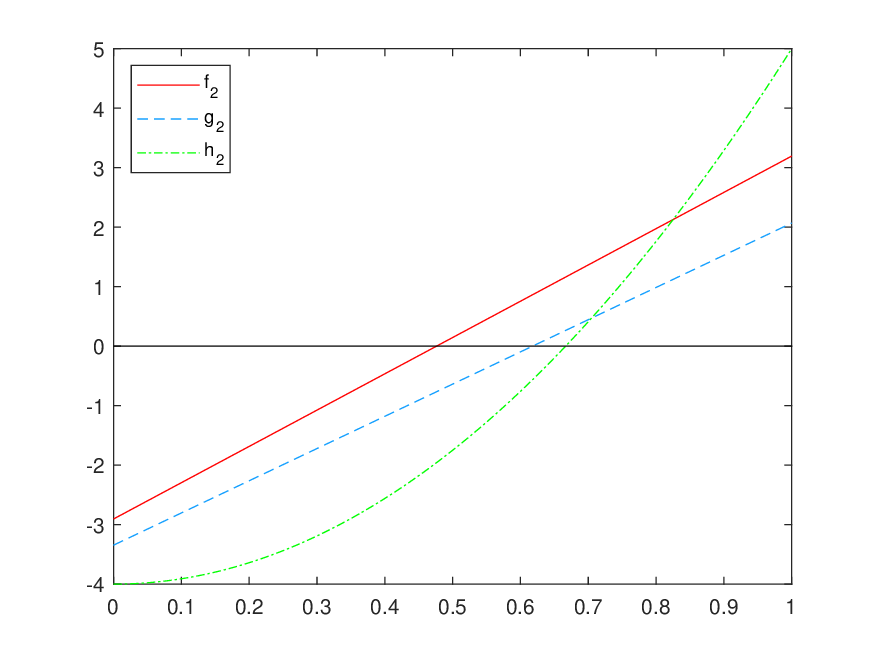}
	\caption{$f_2(x)$ from our result (solid red line), $g_2(x)$ from Theorem 3 in \cite{zhao2022detection} (dash blue line), $h_2(x)$ from Theorem 3 in \cite{Li2019TheNO} (dash-dotted green line).}
	\label{fig:2}
\end{figure}

\textit{\textbf{Example 2}} Consider the four-qubit state $\rho\in H_1^2\otimes H_2^2\otimes H_3^2\otimes H_4^2$,
\begin{equation}\label{32}
\rho=x|\psi\rangle\langle\psi|+\frac{1-x}{16}I_{16},
\end{equation}
where $|\psi\rangle=\frac{1}{\sqrt{2}}(|0000\rangle+|1111\rangle)$, $0\leq x\leq1$,
$I_{16}$ is the $16\times16$ identity matrix. We first consider the detection of bi-separability.
Let $\alpha=\frac{1}{10}$ and $\beta=\frac{6}{5}$, From Theorem 1(i) we set
$f_2(x)=\|F^{l_1|l_2l_3l_4}\|_{tr}-(\frac{1}{10}+\frac{6}{5} \sqrt{\frac{11}{2}})=6.1x-(\frac{1}{10}+\frac{6}{5} \sqrt{\frac{11}{2}})$. $\rho$ is not separable under the bipartition $l_1|l_2l_3l_4$ for $f(x)>0$, i.e., $0.4777<x\leq1$. While according to Theorem 3 in \cite{Li2019TheNO}, $\rho$ is not separable under the bipartition $l_1|l_2l_3l_4$ for $h_2(x)=9x^2-4>0$, i.e., $0.6667<x\leq1$. According to Theorem 3 in \cite{zhao2022detection}, $\rho$ is not separable under the bipartition $l_1|l_2l_3l_4$ for $g_2(x)=(4+\sqrt{2})x-(1+\sqrt{\frac{11}{2}})>0$, i.e., $0.6179<x\leq1$. Fig. \ref{fig:2} shows that our method detects better the bi-separability.

\begin{figure}[H]
	\centering
	\includegraphics[width=0.75\textwidth]{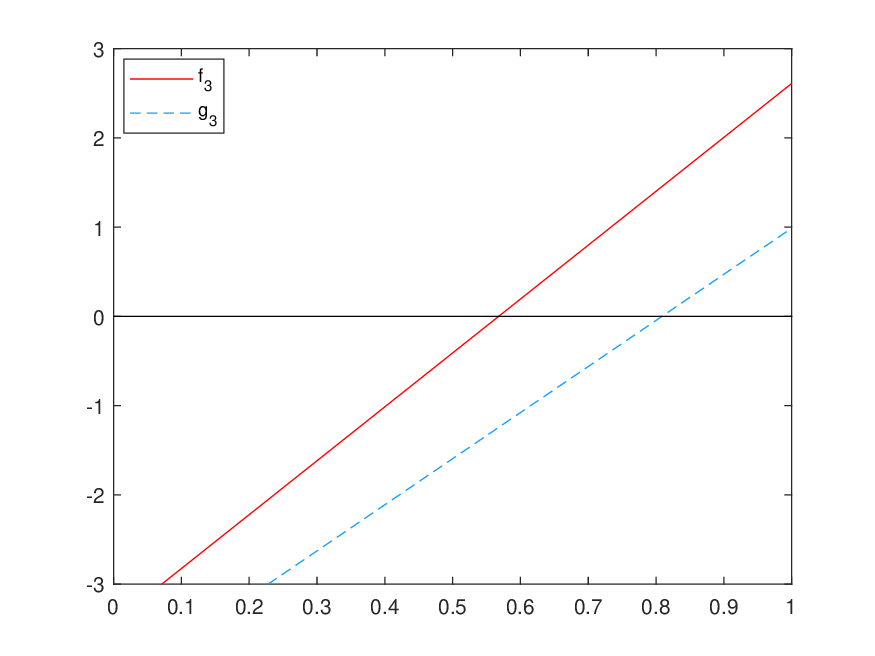}
	\caption{$f_3(x)$ from our result (solid red line), $g_3(x)$ from Corollary 2  in \cite{zhao2022detection} (dashed blue line). }
	\label{fig:3}
\end{figure}

As $\rho$ is a permutational invariant state, by using Corollary 1 we get that $f_3(x)=T(\rho)-J_2=\frac{151}{25}x-\frac{110+12\sqrt{22}+3\sqrt{3}}{50}$
for $\alpha=\frac{1}{10}$ and $\beta=\frac{6}{5}$. $\rho$ is genuine multipartite entangled for $ f_3(x)>0$, i.e., $0.5678<x\leq1$. While according to the Corollary 2 in \cite{zhao2022detection}, $\rho$ is  genuine multipartite entangled for $g_3(x)=T(\rho)-J_2=\frac{23+2\sqrt{2}}{5}x-\frac{11+\sqrt{22}+3\sqrt{3}}{5} \textgreater 0$, i.e., $0.8087<x\leq1$. Fig. \ref{fig:3} shows that our criterion detects better the multipartite entanglement.

\section{Conclusions}
By using the Weyl representation of the generalized Pauli operators, we have presented criteria to certify genuine multipartite entanglement in arbitrary dimensional multipartite quantum systems. This criterion is shown to detect better genuine multipartite entanglement than the existing criteria by detailed examples. The results may shed new light on the applications of genuine multipartite entanglement in quantum information processing. The approach may be also applied to detect other quantum correlations.

\bigskip
\noindent
{\bf Acknowledgements}
This work is supported by National Natural Science
Foundation of China (Grant Nos. 12075159, 12171044);
Beijing Natural Science Foundation (Grant No. Z190005); the Academician
Innovation Platform of Hainan Province.


\end{document}